\documentclass[aps,prd,superscriptaddress,nofootinbib,twocolumn,longbibliography]{revtex4-2}
\usepackage{amsmath,amssymb}
\usepackage{epsfig,graphicx}
\usepackage{xcolor}
\usepackage{natbib}
\usepackage{type1ec}
\usepackage{orcidlink}
\usepackage{bm}

\catcode`@=11
\@addtoreset{equation}{section}
\catcode`@=12

\usepackage{hyperref}
\hypersetup{
    colorlinks,%
    citecolor=blue,%
    filecolor=blue,%
    linkcolor=blue,%
    urlcolor=blue
}
\begin{document}
\title{Spinning test particles in the spacetime of a global monopole}
\author{Morteza Mohseni \orcidlink{0000-0002-5624-0877} }
\email{morteza.mohseni@uleth.ca,\,m-mohseni@pnu.ac.ir}
\affiliation{Theoretical Physics Group, Department of Physics and Astronomy \& Quantum Horizons Alberta, University of Lethbridge, 
4401 University Drive, Lethbridge, Alberta T1K 3M4, Canada}
\affiliation{Physics Department, Payame Noor University, Tehran 19395-3697, Iran}

\author{Saurya Das \orcidlink{0000-0003-1191-8469}}
\email{saurya.das@uleth.ca}
\affiliation{Theoretical Physics Group, Department of Physics and Astronomy \& Quantum Horizons Alberta, University of Lethbridge, 
4401 University Drive, Lethbridge, Alberta T1K 3M4, Canada}

\begin{abstract}
	We investigate the motion of spinning test particles in the spacetime of a global monopole in the framework of
	the Mathisson-Papapetrou-Dixon equations. By making use of the symmetries of the spacetime, we obtain a general exact solution to the
	equations of motion. We show that the particle's trajectories, momenta, and spin can be expressed in terms of three specific functions
	of the polar and azimuthal angles. We also show that the system is completely integrable. 
	We obtain the general non-geodesic trajectory of the particle, and also examine the particular cases
	of radial and planar motion. We compare the non-geodesic trajectories of a spinning particle with a non-spinning particle.
	
\end{abstract}

\maketitle

\section{Introduction}
The field equations of the general theory of relativity determine the local geometry of spacetime, but they leave its global shape,
i.e. its topology, unspecified. On the other hand, the topology of spacetime could lead to important observational effects. An 
example of such effects is the speculated imprints of cosmic topology in the cosmic microwave background anisotropies 
\cite{Cornish_1998}. Thus, investigating spacetime topology has long been of interest \cite{Lachi_ze_Rey_1995}. From this point of 
view, it is interesting to investigate solutions of the Einstein field equations with non-trivial topology. The global monopole 
spacetime is one of such solutions. In fact, it belongs to a class of spacetimes known as topological defects \cite{BRANDENBERGER_1994}. 
As discussed in the latter reference, these defects (cosmic strings, domain walls, global monopoles, and textures) contain energy and 
can seed structure formation. The global monopole space time was obtained in \cite{Barriola_1989} as a spacetime generated by global 
monopoles formed in early universe phase transitions. They can be described by a real scalar field with a Higgs type potential with $O(3)$ 
symmetry. The relevant field equations admit a hedgehog-like solution.

The global monopole spacetime has interesting properties, namely the existence of a solid angle deficit
which breaks the spherical symmetry \cite{Bezerra_de_Mello_2001}, it has a conical singularity \cite{sokolov}, and is
not asymptotically flat. This spacetime and its properties have been studied from various perspectives. To mention a few, the motion 
of test particles around monopoles has been studied in \cite{chakra,banerj,hung,farook}, its topological charge has been
computed in \cite{Tan_2018}, the motion of a relativistic quantum oscillator in a global monopole background has been studied in
\cite{Ahmed_2022}, its possible role in a change in the topology of the universe has been investigated in \cite{MARUNOVIC2016268}, 
global monopole surrounded by quintessence-like matter has been investigated in \cite{Li_2008}, magnetic monopoles in this spacetime 
have been studied in \cite{Spinelly_2002}, the motion of spinning particles in a black hole with a global monopole has been studied in
\cite{hung}, and its general geometric properties have been studied in \cite{shaikh2023geometrical}. Some other aspects of 
these spacetimes have been studied in \cite{barbosa}, see also references therein.

A natural extension of the above-described line of research is a more inclusive study of the motion of particles in the global 
monopole spacetime. This is mainly motivated by the fact that the dynamics of particles is an essential tool in the study of the 
properties of any spacetime. In fact, most of the classical tests of the general theory of relativity and other theories of gravity, 
such as the bending of light or perihelion precession rely on particle dynamics. For test particles this is, roughly speaking,  
equivalent to the study of spacetime geodesics. Particles with internal structures present a different scenario. By treating the 
internal structure by a multipole expansion, and keeping only the pole-dipole terms, a set of equations for the motion of spinning 
test particles has been obtained, usually referred to as the Mathisson-Papapetrou-Dixon (MPD)
equations \cite{1951RSPSA.209..248P,dixon1970}. A main feature of the MPD equations is a spin-curvature coupling which results in
non-geodesic motion, and makes particle's four-momentum and four-velocity being non-parallel in general. 
The only known exception is the trivial case of maximally symmetric spacetimes where the spin-curvature terms vanish leading to 
decoupling of the equations, and geodesic motion. The MPD equations have been subject to extensive studies in various contexts, 
see e.g., \cite{sepangi,tucker,nieto,kyrian,mess,dinesh,bini,puetz,plyt,Mohseni,ramirez,magd,marsot,antonio,fan,yong,pang}.
Geodesic motion of spinning particles in Schwarzschild and Kerr spacetimes has been studied in \cite{nayak} using the 
Ohashi-Kyrian-Semer\'ak spin supplementary condition. It has been shown in \cite{filipe} that exact gravito-electromagnetic analogies 
for the force and spin evolution of spinning test particles arise when the Mathisson-Pirani spin condition is imposed, and this has been used for 
comparing magnetic dipoles in electromagnetic fields with gyroscopes in curved spacetime.

In investigating the MPD equations, researchers have predominantly relied on approximate or numerical methods, driven by both the 
complexity of the equations and the assumption that their resulting deviations from the geodesic motion are minimal. In fact, in 
spite of extensive literature, very few exact solutions to the MPD equations have been known so far, namely the special
solution presented in \cite{Tod_1976}, which describes motion of a spinning particle in the equatorial plane of the Kerr black hole
with the spin vector being perpendicular to this plane. However, the search for exact solutions of the MPD equations is still 
interesting, at least from a formal point of view. For a general spacetime, the difficulty in finding solutions stems from 
the loop-like coupling of the equations, i.e., one has to know the trajectory to determine the momenta and spin components and 
vice versa. 

The aim of the present work is to study the motion of spinning test particles in the spacetime of a global monopole. We find a general
exact solution of the MPD equations for motion in the spacetime of a global monopole. The motivation for this study lies in
both the interest in MPD equations and the global monopole spacetime as a space with topological defect. The significance of our
findings is that, to the best of our knowledge, this is the first general solution to the MPD equations in a non-maximally symmetric
curved spacetime. In the next sections, we first briefly review the global monopole spacetime, and  also the MPD equations. 
We then obtain the explicit form of the MPD equations for the spacetime under consideration and examine their general properties. 
Then, we solve the equations to find the particle's momenta, spin, and trajectories. A Hamiltonian approach is also 
used to show that the system is completely integrable. We consider some physically interesting special 
cases like radial and planar motion. Then, we study the properties of the general solution by considering certain values of the 
relevant parameters. Finally, we discuss the results.


\section{The global monopole spacetime}
The spacetime of a global monopole is described by the following line element
\begin{equation}\label{e1}
	ds^2 = -dt^2+\frac{1}{\alpha^2}dr^2+r^2(d\theta^2+\sin^2\theta d\phi^2)
\end{equation}
in which $\alpha<1$ is a constant so that a sphere of radius $l$ would have an area $4\pi(\alpha l)^2<4\pi l^2$. 
This was first derived in \cite{Barriola_1989} as a solution to the field equation
associated with the Einstein-Hilbert action
\begin{equation}\label{e2}
	S_{EH} = \frac{1}{8\pi G}\int\, {\sqrt{-g}}\,R\, d^4x
\end{equation}
supplemented with the following matter field action
\begin{equation}\label{e3}
	S_{\varphi} = \int{\sqrt{-g}}\left(-\frac{1}{2}\partial_\mu\varphi^a\partial^\mu\varphi^a-V(\varphi^a)\right)\,d^4x
\end{equation}
in which
\begin{equation}\label{e4}
	V(\varphi^a)=\frac{\mu^2}{2}\varphi^a\varphi^a+\frac{\lambda_0}{4}(\varphi^a\varphi^a)^2+\frac{\mu^4}{4\lambda_0}.
\end{equation}
Here, $\mu$ is a mass parameter, and $\lambda_0$ is a coupling constant. The field components $\varphi^a$ with $a=1,2,3$ are scalars,
and the matter field action is $O(3)$ symmetric. The relevant field equations are satisfied by
\begin{equation}\label{e5}
	\varphi^a = \varphi(r) \frac{x^a}{r}
\end{equation}
in which $x^a$ are the Cartesian coordinates $(x,y,z)$.


\section{The MPD equations}
The motion of a spinning test particle in a curved background in the pole-dipole approximation is given by \cite{1951RSPSA.209..248P}
\begin{align}
	\frac{Dp^\mu}{D\tau}\, &=\, -\frac{1}{2}\,{R^\mu}_{\nu\kappa\lambda}\,{\dot x}^\nu\, s^{\kappa\lambda},\label{e6}\\
	\frac{Ds^{\mu\nu}}{D\tau}\, &=\,\, p^\mu\,{\dot x}^\nu - p^\nu\,{\dot x}^\mu,\label{e7}
\end{align}
in which $p^\mu$ is the particle's four-momentum, ${\dot x}^\mu$ is the four-velocity, $s^{\mu\nu}$ is the spin tensor, 
$\frac{D}{D\tau}$ is the absolute derivative, and ${R^\mu}_{\nu\alpha\beta}=\partial_\alpha\Gamma^\mu_{\nu\beta}+\Gamma^\mu_{\alpha\gamma}
\Gamma^\gamma_{\nu\beta}-\alpha\leftrightarrow\beta$ is the curvature tensor. The particle's trajectory is given by ${x^\mu(\tau)}$.
This set of equations is not sufficient for determining all the unknown variables. Thus, they should be supplemented by
some extra equations. Several different such equations have been used in the literature, and the most widely used one is as follows
\cite{dixon1970}
\begin{equation}\label{e8}
	p_\mu s^{\mu\nu} = 0
\end{equation}
which is sometimes called the Tulczyjew condition. The particle momentum and spin are conserved, i.e.,
\begin{align}
	p_\mu p^\mu\, &=\,\, -m^2, \label{e9}\\
	s_{\mu\nu} s^{\mu\nu}\, &=\,\, 2\,s^2. \label{e10}
\end{align}
A more-intuitive alternative way to describe the spin of the particle is to use the spin four-vector defined by
\begin{equation}\label{e11}
	S^\mu = \frac{1}{2\,m\,\sqrt{-g}}\,{\varepsilon^\mu}_{\nu\kappa\lambda}\,p^\nu\, s^{\kappa\lambda},
\end{equation}
in which $\varepsilon^{\mu\nu\kappa\lambda}$ is the alternating symbol with $\varepsilon^{0123}=+1$. It satisfies
\begin{equation}\label{e11aa}
	S_\mu\,S^\mu= s^2
\end{equation}
and
\begin{equation}\label{e8aa}
	p_\mu S^\mu = 0.
\end{equation}

The spinning particle four-velocity and four-momentum are not parallel in general as a result of spin-curvature coupling. It is possible to
obtain an equation for the four-velocity from the above equations of motion, see e.g. \cite{sepangi}. To obtain the particle trajectory,
we need one extra condition (which, in fact, plays the role of the normalization of the four-velocity of test particles). We choose the
following relation for this purpose \cite{sepangi}
\begin{equation}\label{e11a}
	p_\mu\,{\dot x}^\mu\,=\, -m.
\end{equation}
From the MPD equations together with the Tulczyjew condition (Eqs. (\ref{e6})-(\ref{e8})), we obtain
\begin{equation}\label{e11b}
	p^\mu\,-\,m\,{\dot x}^\mu\,=\,-\frac{1}{2m}\,s^{\mu\nu}\,R_{\nu\lambda\alpha\beta}\,{\dot x}^\lambda\,s^{\alpha\beta}.
\end{equation}
For spacetimes admitting a Killing vector ${\pmb\xi}$, the following relation holds \cite{rudiger1,rudiger2}
\begin{equation}\label{killing1}
	p^\mu\,\xi_\mu-\frac{1}{2}\nabla_\nu\,\xi_\mu\,s^{\mu\nu}=\text{const.}
\end{equation}
in which $\nabla$ represents the covariant derivative.

Also, in addition to the above invariants, there are quasi-invariant quantities for backgrounds admitting a
Killing-Yano tensor. Here, quasi-invariant means invariant up to certain orders of quantities like $s^2$.
It has been shown in \cite{rudiger1,rudiger2,druart} that, if $K_{\mu\nu}$ is a Killing-Yano tensor, then the following relation
holds
\begin{equation}\label{killing1a}
	\frac{1}{2\sqrt{-g}}{\varepsilon^{\alpha\beta}}_{\mu\nu}\,K_{\alpha\beta}\,s^{\mu\nu}=\text{const.}+{\mathcal O}(s^2).
\end{equation}
More recent discussions of these conserved quantities may be found in \cite{Ramond}, for motion in Kerr spacetime, and 
in \cite{Andersson}, where massless spinning particles have also been considered.  


\section{General analysis}\label{who}
For a spinning particle moving in a global monopole spacetime, the equation of motion (\ref{e6}) takes the following explicit form
\begin{align}
	\frac{dp^0}{d\tau}\,&=\,0,\label{e12}\\
	\frac{dp^1}{d\tau}\,&=\,\alpha^2\, r\,({\dot\theta}\,p^2+{\sin}^2\theta\,{\dot\phi}\,p^3),\label{e13}\\
	\frac{dp^2}{d\tau}\,&=\,-\frac{{\dot\theta}p^1+{\dot r}p^2}{r}+{\sin}\theta\,{\cos}\theta\,{\dot\phi}\,p^3\nonumber\\
	&\quad -(1-\alpha^2)\,{\sin}^2\theta\,{\dot\phi}\,s^{23},\label{e14}\\
	\frac{dp^3}{d\tau}\,&=\,-\frac{{\dot\phi}p^1+{\dot r}p^3}{r}-{\cot}\theta\,{\dot\theta}\,p^3\nonumber\\&\quad
	-\cot\theta\,{\dot\phi}\,p^2+(1-\alpha^2){\dot\theta}\,s^{23},\label{e15}
\end{align}
while Eq. (\ref{e7}) can be expanded as
\begin{align}
	\frac{ds^{01}}{d\tau}\,&=\,\alpha^2\,r\,({\dot\theta}\,s^{02}+{\sin}^2\theta\,{\dot\phi}\,s^{03})
	+p^0{\dot r}-p^1{\dot t},\label{e16}\\
	\frac{ds^{02}}{d\tau}\,&=\,-\frac{{\dot r}\,s^{02}+{\dot\theta}\,s^{01}}{r}+\sin\theta\,\cos\theta\,{\dot\phi}\,s^{03}\nonumber\\&\quad
	+p^0\,{\dot\theta}-p^2\,{\dot t},\label{e17}\\
	\frac{ds^{03}}{d\tau}\,&=\,-\frac{{\dot r}\,s^{03}+{\dot\phi}\,s^{01}}{r}-\cot\theta\,({\dot\phi}\,s^{02}+{\dot\theta}\,s^{03})\nonumber
	\\&\quad +p^0\,{\dot\phi}-p^3\,{\dot t},\label{e18}\\
	\frac{ds^{12}}{d\tau}\,&=\,-\alpha^2\,r\,\sin^2\theta\,{\dot\phi}\,s^{23}+\sin\theta\,\cos\theta\,{\dot\phi}\,s^{13}-
	\frac{\dot r}{r}\,s^{12}\nonumber\\&\quad +p^1\,{\dot\theta}-p^2\,{\dot r},\label{e19}\\
	\frac{ds^{13}}{d\tau}\,&=\,-\cot\theta\,({\dot\theta}\,s^{13}+{\dot\phi}\,s^{12})-\frac{\dot r}{r}\,s^{13}
	+\alpha^2\,r\,{\dot\theta}\,s^{23}\nonumber\\&\quad +p^1\,{\dot\phi}-p^3\,{\dot r},\label{e20}\\
	\frac{ds^{23}}{d\tau}\,&=\,-\cot\theta\,{\dot\theta}\,s^{23}-\frac{{\dot\theta}\,s^{13}-{\dot\phi}\,s^{12}+2{\dot r}\,s^{23}}{r}
	\nonumber\\&\quad +p^2\,{\dot\phi}-p^3\,{\dot\theta},\label{e21}
\end{align}
and Eq. (\ref{e8}) becomes
\begin{align}
	p_1\,s^{01}+p_2\,s^{02}+p_3\,s^{03}\,&=\,0,\label{e22}\\
	p_0\,s^{01}-p_2\,s^{12}-p_3\,s^{13}\,&=\,0,\label{e23}\\
	p_0\,s^{02}+p_1\,s^{12}-p_3\,s^{23}\,&=\,0,\label{e24}\\
	p_0\,s^{03}+p_1\,s^{13}+p_2\,s^{23}\,&=\,0.\label{e25}
\end{align}

It would also be useful to obtain the explicit form of Eq. (\ref{e11b}). It is
\begin{equation}\label{e11c}
	p^\mu\,-\,m\,{\dot x}^\mu\,=\,\frac{1-\alpha^2}{m}\,r^2\,\sin^2\theta\,s^{23}\,({\dot\theta}\,s^{\mu 3}-{\dot\phi}\,s^{\mu 2})
\end{equation}
The right hand side of the above equation shows that the spin-curvature coupling is proportional to $s^{23}$. Obviously, it vanishes
in the Minkowski spacetime, for which $\alpha=1$. For $\alpha\neq 1$, it only vanishes when the motion is radial,
$\dot\theta=\dot\phi=0$, or $s^{23}=0$. In terms of components, we have
\begin{align}
	p^0\,-\,m\,{\dot t}\,&=\,\Psi\,({\dot\theta}\,s^{03}-{\dot\phi}\,s^{02}),\label{e11d}\\
	p^1\,-\,m\,{\dot r}\,&=\,\Psi\,({\dot\theta}\,s^{13}-{\dot\phi}\,s^{12}),\label{e11e}\\
	p^2\,-\,m\,{\dot\theta}\,&=\,\Psi\,{\dot\theta}\,s^{23},\label{e11f}\\
	p^3\,-\,m\,{\dot\phi}\,&=\,\Psi\,{\dot\phi}\,s^{23},\label{e11g}
\end{align}
in which
\begin{equation}\label{psi}
	\Psi\,=\,\frac{1-\alpha^2}{m}\,r^2\,\sin^2\theta\,s^{23}.
\end{equation}
Now, Eq. (\ref{e11f}) implies that for ${\dot\theta}\,=\,0$, we have $p^2\,=\,0$. Then, by inserting $\theta\,=\,\frac{\pi}{2}$ into 
Eq. (\ref{e14}), we get $s^{23}\,=\,0$ and hence $\Psi=0$, which in turn means that momentum and velocity are parallel. Thus, planar motion 
is possible only when spin of the particle does not couple to the monopole spacetime curvature.

It can be shown that the global monopole spacetime admits the following Killing-Yano tensor
\begin{equation}\label{k13}
	\sigma_{\mu\nu}=\left\{
	\begin{array}{ll}
		r^3\,\sin\theta  & \hspace{5mm}\mu=2,\nu=3       \\
		-r^3\,\sin\theta & \hspace{5mm}\mu=3,\nu=2       \\
		0                & \hspace{5mm}\mbox{otherwise}.
	\end{array}
	\right.
\end{equation}
It is also a Killing-Yano tensor for the spherically symmetric spacetimes considered in \cite{acik,horwat}.
Inserting this into Eq. (\ref{killing1a}) results in
\begin{equation}\label{k14}
	s^{01}\,=\,\frac{L}{r}+{\mathcal O}(s^2)
\end{equation}
where $L$ is constant. In fact, it can be shown from the equations of motion that the following relation holds
\begin{equation}\label{app}
	\frac{d}{d\tau}(r\,s^{01})=\frac{1}{2\,E}\,\frac{d}{d\tau}(s_{12}\,s^{12}+s_{13}\,s^{13}+s_{23}\,s^{23}).
\end{equation}
While the above relation might be useful in future investigations, in the present work, we use exact expressions for $s^{01}$.


\section{General motion}\label{genmot}
In this section, we turn to finding the general solution to the equations of motion of the spinning particle in the spacetime of a global 
monopole. From Eq. (\ref{e12}) we obtain
\begin{equation}\label{g1}
	p^0\,=\,E
\end{equation}
in which $E$ is a constant. Also, by combining Eqs. (\ref{e13}) and (\ref{e11a}), we obtain
\begin{equation}\label{g2}
	p^1\,=\,\frac{\alpha^2}{r}\,(E\,t-m\,\tau)+\frac{r_0\beta}{r}
\end{equation}
in which $r_0$ is the initial value of $r$ and $\beta$ is the value of $p_1$ at $\tau=0$. Note that we have assumed that
$t(\tau=0)=0$. We expect that for $\alpha=1$, we should have $p^2=p^3=0$, hence, from Eq. (\ref{e9}),
\begin{equation}\label{beta1}
	\beta=\pm\alpha\sqrt{E^2-m^2}.
\end{equation}
For a particle at rest, $\beta=0$. Note that Eq. (\ref{g2}) is not in fact an
explicit solution for $p^1$ as it is expressed in terms of $t$ whose dependence on $\tau$ is to be determined.

For the spacetime described by the metric given in Eq. (\ref{e1}), it can be shown that the following four-vectors satisfy the
Killing equation
\begin{align}
	\xi^\mu\,&=\,(1,0,0,0),\label{k1}\\
	\zeta^\mu\,&=\,(0,0,0,1),\label{k2}\\
	\chi^\mu\,&=\,(0,0,-\cos\phi,\cot\theta\,\sin\phi),\label{k3}\\
	\rho^\mu\,&=\,(0,0,\sin\phi,\cot\theta\,\cos\phi).\label{k4}
\end{align}
Inserting Eq. (\ref{k1}) into Eq. (\ref{killing1}) returns Eq. (\ref{g1}), serving as a consistency check. Now, by inserting Eq. (\ref{k2}) 
into Eq. (\ref{killing1}) we obtain
\begin{equation}\label{k7}
	r\,\sin^2\theta\,(r\,p^3+s^{13})+r^2\,\sin\theta\,\cos\theta\,s^{23}\,=\,q_1
\end{equation}
in which $q_1$ is constant. Similarly, by inserting Eqs. (\ref{k3}) and (\ref{k4}) into Eq. (\ref{killing1}), we obtain
\begin{align}
	-r\,\cos\phi\,(r\,p^2+s^{12})\,&+\,r\,\sin\theta\,\cos\theta\,\sin\phi\,(r\,p^3+s^{13})\nonumber\\&-\,r^2\,
	\sin^2\theta\,\sin\phi\,s^{23}\,=\,q_2\label{k8},\\
	r\,\sin\phi\,(r\,p^2+s^{12})\,&+\,r\,\sin\theta\,\cos\theta\,\cos\phi\,(r\,p^3+s^{13})\nonumber\\&-\,r^2\,
	\sin^2\theta\,\cos\phi\,s^{23}\,=\,q_3\label{k9},
\end{align}
respectively. Here, $q_2$ and $q_3$ are constants. The constants $q_{1,2,3}$ can be related to the three independent orientations of
the spin vector. The algebraic equations (\ref{k7})-(\ref{k9}) result in
\begin{align}
	r\,s^{12} \,&=\, Q_1-r^2p^2\label{k10},\\
	r\,\sin\theta\, s^{13}\, &=\, Q_2-r^2\sin\theta\, p^3\label{k11},\\
	r^2\,\sin\theta\, s^{23}\, &=\, Q_3,\label{k12}
\end{align}
in which
\begin{align}
	Q_1\,&=\,q_3\,\sin\phi-q_2\,\cos\phi\label{k10a},\\
	Q_2\,&=\,q_1\,\sin\theta+\cos\theta\,(q_2\,\sin\phi+q_3\,\cos\phi)\label{k11a},\\
	Q_3\,&=\,q_1\,\cos\theta-\sin\theta\,(q_2\,\sin\phi+q_3\,\cos\phi).\label{k12a}
\end{align}
We also have the following useful relations
\begin{equation}\label{u1}
	Q_1^2+Q_2^2+Q_3^2\,=\,q^2,
\end{equation}
where
\begin{equation}\label{qqq}
q^2\equiv q_1^2+q_2^2+q_3^2
\end{equation}
and
\begin{align}
	{\dot Q}_1 \,&=\, (Q_2\cos\theta-Q_3\sin\theta)\,{\dot\phi},\label{u2}\\
	{\dot Q}_2 \,&=\, Q_3\,{\dot\theta}-Q_1\,\cos\theta\,{\dot\phi},\label{u3}\\
	{\dot Q}_3 \,&=\, -Q_2\,{\dot\theta}\,+Q_1\,\sin\theta\,{\dot\phi},\label{u4}
\end{align}
where a dot over a variable means $\frac{d}{d\tau}$. 
Now, we can insert Eqs. (\ref{k10})-(\ref{k12}) into Eqs. (\ref{e23})-(\ref{e25}) to obtain
\begin{align}
	s^{01}\,&=\,\frac{1}{E}\left(r^3\,(p^2)^2+r^3\,\sin^2\theta\,(p^3)^2\right.\nonumber\\&\quad\left.
	-Q_1\,r\,p^2-Q_2\,r\,\sin\theta\,p^3\right),\label{k16}\\
	s^{02}\,&=\,\frac{1}{E}\left(\frac{Q_1\,p^1}{r\,\alpha^2}-\frac{p^1}{\alpha^2}\,r\,p^2-\sin\theta\,
	Q_3\,p^3\right),\label{k17}\\
	s^{03}\,&=\,\frac{1}{E}\left(\frac{Q_2\,p^1}{r\,\sin\theta\,\alpha^2}-\frac{p^1}{\alpha^2}\,r\,p^3+
	\frac{Q_3}{\sin\theta}\,p^2\right),\label{k18}
\end{align}
respectively. Alternatively, in terms of the spin vector defined in (\ref{e11}), we have
\begin{align}
	S^0\,&=\,\frac{1}{mE\alpha}\,(E\,\Sigma),\label{sp1}\\
	\frac{S^1}{\alpha}\,&=\,\frac{1}{mE\alpha}\,\left(m^2\,\alpha\, Q_3+\frac{p^1}{\alpha}\,\Sigma\right),\label{sp2}\\
	r S^2\,&=\,\frac{1}{mE\alpha}\,\left(-m^2\, r\,\sin\theta\, s^{13}+r\,p^2\,\Sigma\right),\label{sp3}\\
	r\sin\theta S^3\,&=\,\frac{1}{mE\alpha}\left(m^2\, r\,s^{12}+r\,\sin\theta\, p^3\,\Sigma\right),\label{sp4}
\end{align}
in which
\begin{equation}
	\Sigma=Q_3\,p^1-Q_2\,r\,p^2+Q_1\,r\,\sin\theta\,p^3\label{je2}
\end{equation}
and $p^1, s^{12}, s^{13}$ are given by Eqs. (\ref{g2}), (\ref{k10}), (\ref{k11}), respectively.

Thus, up to now, we have solved the relevant equations to obtain $p^0,s^{23}$ completely, $p^1$ in terms of $t$, and
$s^{01},s^{02},s^{03},s^{12},s^{13}$ in terms of $p^2,p^3$. The above relations, expressing the dynamical variables in
terms of $p^2,p^3$, also typically depend on $r$ which is itself unknown unless we already know the trajectory. Thus, to
determine the dynamical variables, we need some more steps.
We start by inserting Eq. (\ref{k12}) into Eq. (\ref{psi}) to arrive at
\begin{equation}\label{psi1}
	\Psi = \frac{1-\alpha^2}{m}\,Q_3\,\sin\theta.
\end{equation}
Then, Eqs. (\ref{e11f}) and ({\ref{e11g}}) reduce to
\begin{align}
	{\dot\theta}\,&=\,\frac{p^2}{M},\label{e91f}\\
	{\dot\phi}\,&=\,\frac{p^3}{M},\label{e91g}
\end{align}
in which
\begin{equation}\label{e912d}
	M=m+\frac{(1-\alpha^2)\,Q_3^2}{m\,r^2}.
\end{equation}
We can also obtain a relation between $p^1$ and ${\dot r}$ by deploying Eqs. (\ref{e91f}), (\ref{e91g}), (\ref{k10}), (\ref{k11}), 
(\ref{psi}), and (\ref{e11e}). This results in
\begin{equation}\label{e91e}
	{\dot r} = \frac{1}{m}\,\left(p^1+\frac{(1-\alpha^2)\,Q_3\,{\dot Q}_3}{m\,r}\right).
\end{equation}
For convenience, let us summarize what we have done in this section so far. We have directly solved the equations of 
motion to obtain $p^0$ and (implicitly) $p^1$. The relevant expressions involve the constants $\alpha, \beta, E$, and of course, $m$. We have also 
used a set of Killing vectors $\zeta^\mu,\chi^\mu,\rho^\mu$ to obtain three constants of motion $q_{1,2,3}$. The resulting expressions have been 
simplified by introducing three functions $Q_{1,2,3}(\theta,\phi)$ which have convenient properties. The function $Q_3$ essentially gives $S^{23}$, 
$Q_1$ gives $S^{12}$ in terms of $p^2$, and $Q_2$ gives $S^{13}$ in terms of $p^3$. The remaining spin components $S^{01,02,03}$ have been
obtained using the supplementary condition. Finally, ${\dot r},{\dot\theta}$, and ${\dot\phi}$ have been related to $p^1,p^2$, and $p^3$, respectively.  
To make the resulting expressions more compact, we have also introduced the abbreviations $q, \Psi, \Sigma$, and $M$.

To proceed further, we first rewrite Eqs. (\ref{e14}) and (\ref{e15}) into the following forms
\begin{align}
	\frac{d(r^2\,p^2)}{d\tau}\,&=\,-{\dot\theta}\,r\,p^1+r\,{\dot r}\,p^2+\cos\theta\,{\dot\phi}\,(r^2\,\sin\theta\,p^3)\nonumber\\
	&\quad-(1-\alpha^2)\,\sin\theta\,{\dot\phi}\,Q_3,\label{e1400}\\
	\frac{d(r^2\,\sin\theta\,p^3)}{d\tau}\,&=\,-{\dot\phi}\,r\sin\theta\,p^1+r\,{\dot r}\,\sin\theta\,p^3\nonumber\\&\quad
	-\cos\theta\,{\dot\phi}\,(r\,p^2)+(1-\alpha^2){\dot\theta}\,Q_3,\label{e1500}
\end{align}
respectively. Now, if we assume that neither ${\dot\theta}$ nor $\dot\phi$ vanish identically (as in the case of radial motion), 
the second term in the right-hand side of Eq. (\ref{e1400}) can be rewritten as
\begin{align}
	r\,{\dot r}\,p^2\, &=\, r\,{\dot r}\, M {\dot\theta}\nonumber\\
	\,&=\, r\,{\dot\theta}\,\left(p^1+\frac{(1-\alpha^2)\Sigma}{m^2 r^2}Q_3\right),\label{vv1}
\end{align}
in which Eqs. (\ref{u4}), (\ref{je2}), and (\ref{e91f})-(\ref{e91e}) have been used. Similarly, the second term
on the right-hand side of Eq. (\ref{e1500}) can be rewritten as
\begin{equation}
	r\,{\dot r}\,\sin\theta\, p^3 = r\,{\dot\phi}\,\sin\theta\,\left(p^1+\frac{(1-\alpha^2)\Sigma}{m^2 r^2}Q_3\right).\label{vv2}
\end{equation}
Thus, Eqs. (\ref{e1400}) and (\ref{e1500}) reduce to 
\begin{align}
	\frac{d(r^2\,p^2)}{d\tau}\,&=\frac{(1-\alpha^2)\Sigma}{m^2 r}\,Q_3\,{\dot\theta}+\cos\theta\,{\dot\phi}\,(r^2\,\sin\theta\,p^3)\nonumber\\
	&\quad-(1-\alpha^2)\,\sin\theta\,{\dot\phi}\,Q_3,\label{e1401}\\
	\frac{d(r^2\,\sin\theta\,p^3)}{d\tau}\,&=\frac{(1-\alpha^2)\Sigma}{m^2 r}\,Q_3\,{\dot\phi}\,\sin\theta\nonumber\\&\quad
	-\cos\theta\,{\dot\phi}\,(r\,p^2)+(1-\alpha^2){\dot\theta}\,Q_3,\label{e1501}
\end{align}
where Eq. (\ref{k12}) has been used.

On the other hand, from Eq. (\ref{k10}), we can infer the following structure 
\begin{equation}
	r^2\,p^2 \,=\, (1-\alpha^2)Q_1+F,\label{ansz1}
\end{equation}
which is justified as follows: in the absence of spin, we can set $F=0$, then both Eqs. (\ref{k10}) and (\ref{ansz1}) consistently give
$r^2p^2\propto Q_1$. The factor $1-\alpha^2$ in Eq. (\ref{ansz1}) ensures that for flat spacetime $p^2$ vanishes. By similar arguments, we can
write 
\begin{equation}
	r^2\,\sin\theta\, p^3\, =\, (1-\alpha^2)Q_2+G.\label{ansz2}
\end{equation}
In the latter two equations $F,G$ are functions to be determined.

Inserting Eqs. (\ref{ansz1}) and (\ref{ansz2}) into Eqs. (\ref{e1401}) and (\ref{e1501}), and making use of Eqs. (\ref{u2})-(\ref{u4}),
we obtain  
\begin{align}
	{\dot F}\, &=\, {\dot\theta}\,\frac{(1-\alpha^2)Q_3}{m^2 r}\Sigma +{\dot\phi}\,\cos\theta\, G,\label{ansz5}\\
	{\dot G}\, &=\, {\dot\phi}\left(-F\cos\theta+\frac{(1-\alpha^2)Q_3}{m^2 r}\,\Sigma\,\sin\theta\right).\label{ansz6}
\end{align}
Comparing these relations with Eqs. (\ref{u2})-(\ref{u4}) we obtain the following solutions
\begin{align}
	F \,&=\, \lambda\, Q_2,\label{ansz7}\\
	G \,&=\, -\lambda\, Q_1,\label{ansz8}\\
	(1-\alpha^2)\Sigma\,&=\,\lambda\, m^2\, r,\label{ansz9}
\end{align}
in which $\lambda$ is a constant. Finally, by inserting Eqs. (\ref{ansz7}) and (\ref{ansz8}) back into Eqs. (\ref{ansz1})
and (\ref{ansz2}), we obtain
\begin{align}
	r^2\,p^2 \,&=\, (1-\alpha^2)Q_1+\lambda Q_2,\label{ansz10}\\
	r^2\, \sin\theta\, p^3 \,&=\, (1-\alpha^2)Q_2-\lambda Q_1.\label{ansz11}
\end{align}
Also, from Eqs. (\ref{ansz10}), (\ref{ansz11}), and (\ref{e9}) we get
\begin{equation}\label{k20a}
	r\,p^1\,=\,\pm\sqrt{\beta^2\,r^2-\alpha^2(\lambda^2+(1-\alpha^2)^2)\,{\mathcal Q}},
\end{equation}
in which $\beta$ is defined by Eq. (\ref{beta1}), and ${\mathcal Q}\equiv Q_1^2+Q_2^2$. This completes finding solutions for the 
momentum components. 

The components of spin four-vector can be obtained by first inserting Eqs. (\ref{ansz10}) and (\ref{ansz11}) into Eqs. (\ref{k10}) and 
(\ref{k11}) to find $rs^{12}$ and $r\sin\theta s^{13}$, and then using the resulting expressions together with Eqs. (\ref{ansz9})
and (\ref{sp1})-(\ref{sp4}) to get
\begin{align}
	S^0\, &=\, \frac{m\lambda}{\alpha(1-\alpha^2)}\,r,\label{sp1100}\\
	S^1\, &=\, \frac{m}{E}\,\left(\alpha Q_3+\frac{\lambda}{\alpha(1-\alpha^2)}\,r\,p^1\right),\label{sp2100}\\
	S^2\, &=\, -\frac{m}{E\alpha r}\,\left(\alpha^2-\frac{\lambda^2}{1-\alpha^2}\right)Q_2,\label{sp3100}\\
	S^3\, &=\, \frac{m}{E\alpha r\sin\theta}\left(\alpha^2-\frac{\lambda^2}{1-\alpha^2}\right)Q_1.\label{sp4100}
\end{align}

Thus, we have determined $p^1,p^2,p^3, S^0, S^1, S^2, S^3$ as functions of $r,\theta,\phi$. These can be 
used to obtain the particle's worldlines. We can still proceed further by expressing $r$
as a function of $\theta,\phi$. To achieve this, we first note that by inserting Eqs. (\ref{ansz10}) and (\ref{ansz11}) into Eq. (\ref{je2})
we obtain
\begin{equation}\label{k200}
	r\,\Sigma = Q_3\, r\, p^1 - \lambda\,{\mathcal Q}.
\end{equation}
Now, by inserting Eqs. (\ref{sp1100})-(\ref{sp4100}) into Eq. (\ref{e11aa}) and making use of Eqs. (\ref{e9}), (\ref{u1}), (\ref{ansz9}),
and (\ref{k200}) we obtain
\begin{align}
	r^2 &= \,\left(\frac{1-\alpha^2}{m\lambda}\right)^2\left\{\left(\frac{Es\alpha}{m}\right)^2- \alpha^2 q^2\right.\nonumber\\
	&\quad +(\alpha^2-\alpha^4-\lambda^2){\mathcal Q}\Bigg\}.\label{e61}
\end{align}

Now, we can insert Eqs. (\ref{ansz10}), (\ref{ansz11}), and (\ref{e61}) into Eqs. (\ref{e91f})-(\ref{e912d}) to obtain
\begin{align}
	{\dot\theta}\, &=\,\frac{m\,((1-\alpha^2)Q_1+\lambda Q_2)}{m^2r^2+(1-\alpha^2)Q_3^2},\label{pl1}\\
	{\dot\phi} \,&=\, \frac{m\,((1-\alpha^2)Q_2-\lambda Q_1)}{m^2r^2+(1-\alpha^2)Q_3^2}.\label{pl2}
\end{align}
By integrating these two equations, we obtain $\theta(\tau), \phi(\tau)$, which can be inserted back into Eq. (\ref{e61}) to obtain $r(\tau)$,
and hence the particle's trajectory is completely determined. Then, by inserting $r(\tau),\theta(\tau),\phi(\tau)$ into
Eqs. (\ref{ansz10}), (\ref{ansz11}), (\ref{k20a}), and (\ref{sp1100})-(\ref{sp4100}) the particle's momentum and spin are also obtained.
Thus, we have a complete set of solutions for all the dynamical variables.

As an interesting consistency check, one may calculate $s^{01}$ by using Eq. (\ref{k16}) to obtain 
\begin{equation}\label{k16000}
  s^{01} = (\lambda^2-\alpha^2(1-\alpha^2))\left(\frac{q^2}{Er}-\frac{Q^2_3}{Er}\right)
\end{equation}	
where the term $\frac{Q^2_3}{Er}\sim\frac{s^2}{Er}$ is quadratic in spin. Thus, we have $s^{01}=\frac{const.}{r}+{\mathcal O}(s^2)$, which is 
the same as Eq. (\ref{k14}). As described earlier in Sec. \ref{who}, it arises due to the existence of the Killing-Yano tensor Eq.(\ref{k13}).

Also, an interesting question is whether the above solution remains valid if one uses a different supplementary condition. In fact, 
the exact solution we found relies on the existence of the Killing vectors given by Eqs. (\ref{k1})-(\ref{k4}) which are properties of the background 
spacetime, the constants of motion given by Eq. (\ref{killing1}), and Eqs. (\ref{e9}) and (\ref{e10}), which still hold at least for the 
Mathisson-Pirani supplementary condition. Thus, it seems the machinery developed here still works, although the specific form of the exact 
solutions might be different. This might not be the case for other supplementary conditions, namely, the Newton-Wigner condition 
(see, e.g., \cite{newton}), the Corinaldesi-Papapetrou condition, and the Ohashi-Kyrian-Semer\'ak condition (see \cite{OtherSSC} for a review of the latter 
two conditions). Comparison of these supplementary conditions in other contexts may be found in \cite{wormhole,kerr}. It is argued in \cite{Andersson} 
that the generalized Carter's constant, as defined there, is conserved independently of which supplementary condition is chosen.  
An elaboration of this issue is left for a future publication.

In addition, the chaotic motion of spinning particles associated with some spherically symmetric spacetimes \cite{chaos} 
is not seen here due to the existence of the above-mentioned constants of motion. 

In Sec. \ref{geodesy}, we briefly go through a few specific cases, and then we will address the general case in Sec. \ref{nogeodesy}.


\section{Complete Integrability}
From Eq. (\ref{k16000}) we can derive the following constant of motion
\begin{equation}\label{k16000a}
rs^{01} + \frac{(\lambda^2-\alpha^2(1-\alpha^2))}{E}r^4\sin^2\theta(s^{23})^2 = J,
\end{equation}	
where $J=\frac{(\lambda^2-\alpha^2(1-\alpha^2))q^2}{E}$.

The Hamiltonian for the spinning particle with the Tulczyjew supplementary condition and the parametrization described by Eq. (\ref{e11a}) is
\begin{equation}\label{H1}
  H = \frac{1}{2m}\left(g^{\mu\nu}+\frac{4{R^\mu}_{\alpha\beta\gamma}s^{\alpha\nu}s^{\beta\gamma}}{R_{\alpha\beta\gamma\delta}s^{\alpha\beta}
  s^{\gamma\delta}-4p_\alpha p^\alpha}\right)p_\mu p_\nu,
\end{equation}	
see Eq. (C.6) of \cite{witznay}. By using Eq. (\ref{k12}), we can simplify the above Hamiltonian to
\begin{equation}\label{H2}
  H = \frac{1}{2m}g^{\mu\nu}p_\mu p_\nu+\frac{(1-\alpha^2)Q_3\sin\theta}{Mm}(p^2s^{3\nu}+p^3s^{2\nu})p_\nu
\end{equation}	
where $M$ is defined by Eq. (\ref{e912d}). By using Eqs. (\ref{e9}) and (\ref{k12}), we obtain $H\cong -m$ in which $\cong$ means the relation
holds only on-shell. 

The Poisson brackets are given by \cite{witznay,witznay2}
\begin{align}
\{x^\mu,x^\nu\}&\, =\,  0, \label{H3}\\
\{x^\mu,p_\nu\}& \,=\,  \delta^\mu_\nu, \label{H4}\\
\{p_\mu,p_\nu\}& \,=\,  -\frac{1}{2}R_{\mu\nu\alpha\beta}s^{\alpha\beta}, \label{H5}\\
\{s^{\mu\nu},p_\kappa\}& \,=\,  -\left(\Gamma^\mu_{\alpha\kappa}s^{\kappa\nu}+\Gamma^\nu_{\alpha\kappa}s^{\mu\kappa}\right), \label{H6}\\
\{s^{\mu\nu},x^\kappa\}& \,= \, 0, \label{H7}\\
\{s^{\mu\nu},s^{\alpha\beta}\}& \,=\,  g^{\mu\alpha}s^{\nu\beta}-g^{\mu\beta}s^{\nu\alpha}+g^{\nu\beta}s^{\mu\alpha}
-g^{\nu\alpha}s^{\mu\beta}. \label{H8}
\end{align} 
The system admits the set of integrals $\{\mathcal C\}_n$, in which $n=1,\cdots,7$, and
\begin{align}
{\mathcal C}_1& =  p_0, \label{H9}\\
{\mathcal C}_2 & =  g^{\alpha\beta}p_\alpha p_\beta, \label{H10}\\
{\mathcal C}_3& = g_{\mu\alpha}g_{\nu\beta}s^{\mu\nu}s^{\alpha\beta}, \label{H11}\\
{\mathcal C}_4& =  r\,\sin^2\theta\,(r\,p^3+s^{13})+r^2\,\sin\theta\,\cos\theta\,s^{23},\label{H12}\\
{\mathcal C}_5& =  -r\,\cos\phi\,(r\,p^2+s^{12})+r\,\sin\theta\,\cos\theta\,\sin\phi(r\,p^3+s^{13})\nonumber\\&\quad-r^2\,
	\sin^2\theta\,\sin\phi\,s^{23}, \label{H13}\\
{\mathcal C}_6& =  r\,\sin\phi\,(r\,p^2+s^{12})+\,r\,\sin\theta\,\cos\theta\,\cos\phi\,(r\,p^3+s^{13})\nonumber\\&\quad-r^2\,
	\sin^2\theta\,\cos\phi\,s^{23}, \label{H14}\\
{\mathcal C}_7& =  rs^{01} + \frac{(\lambda^2-\alpha^2(1-\alpha^2))}{E}\, r^4\sin^2\theta(s^{23})^2. \label{H15}
\end{align} 
These integrals correspond to Eqs. (\ref{g1}), (\ref{e9}), (\ref{e10}), (\ref{k7})-(\ref{k9}), and (\ref{k16000a}), respectively. 
The integrals ${\mathcal C}_{2,3}$ represent generic constants of motions of the MPD equations while ${\mathcal C}_{1,4,5,6}$ originate
    from the explicit symmetries of the background spacetime. The integral ${\mathcal C}_7$, on the other hand, has a rather different nature. This purely 
    spin-dependent constant arises from the hidden symmetry of the global monopole spacetime associated with the Killing-Yano tensor given in 
    Eq. (\ref{k13}). Here, ${\mathcal C}_7$ is an exact conserved quantity containing quadratic spin terms, contrary to the common practice in the literature, 
	where quadratic terms are neglected and Killing-Yano tensors are associated with only quasi-conserved quantities.  

The number of integrals is equal to the number of degrees of freedom $(t,r,\theta,\phi,s^{12},s^{13},s^{23})$. They also satisfy 
\begin{equation}\label{H16}
	\{{\mathcal C}_n,H\}\,=\, 0,\,\,\, n=1,\cdots 7.
\end{equation}
Now, we can also check that the following relations hold
\begin{equation}\label{H17}
   \{{\mathcal C}_n,{\mathcal C}_{n^\prime}\}\,=\,0,\,\,\, n,n^\prime=1,\cdots,7.
\end{equation}	
The calculations resulting in the above relations are lengthy but straightforward. For instance, for $n=2,n^\prime=7$, it is as follows. 
From Eqs. (\ref{H10}) and (\ref{H15}) we have
\begin{align*}
\{{\mathcal C}_2,{\mathcal C}_7\} &= \{g^{\mu\nu}p_\mu p_\nu, r s^{01}\}\\ &+
\frac{(\lambda^2-\alpha^2(1-\alpha^2))}{E}\{g^{\mu\nu}p_\mu p_\nu, r^4\sin^2\theta (s^{23})^2\}\\ &=
-2\alpha^2 p_\mu s^{0\mu}\\&+ \frac{(\lambda^2-\alpha^2(1-\alpha^2))}{E\,r}\,s_{23}(p^2 s^{13} - p^3 s^{12}), 
\end{align*}	
in which Eqs. (\ref{H3})-(\ref{H8}) have been used. Now, the first term vanishes on using Eq. (\ref{e8}). The second term also vanishes by 
using the expressions for $p^2, p^3, s^{12}, s^{13}$ from Eqs. (\ref{ansz10}), (\ref{ansz11}), (\ref{k10}), and (\ref{k11}), respectively. 
Thus, we obtain $\{{\mathcal C}_2,{\mathcal C}_7\}=0$.

Thus, the system is completely (Liouville) integrable. This is reflected in Eqs. (\ref{pl1}) and (\ref{pl2}) (using Eq. (\ref{e61})) where the radial 
and angular parts are decoupled. The symmetries of the spacetime, including the hidden symmetry generated by the Killing-Yano tensor, are 
responsible for integrability of the equations of motion.  A general review of the complete integrability can be found in Appendix B of \cite{frolov}.


\section{Geodesic motion}\label{geodesy}
As we discussed earlier in Sec. \ref{who}, there are situations where the spin-curvature coupling vanishes resulting in $p^\mu$
and ${\dot x}^\mu$ being parallel, and the particle moves along a geodesic. Before embarking on a little more detailed study of a few 
instances of such situations in this section, a closer look into the nature of the general solution presented in the preceding section 
would be illuminating. For the time being, let us decompose Eqs. (\ref{k10}) and (\ref{k11}) into the following form: 
$rs^{12}=\alpha^2 Q_1$, $r\sin\theta s^{13}=\alpha^2 Q_2$, $r^2p^2=(1-\alpha^2)Q_1$,$r^2\sin\theta p^3=(1-\alpha^2)Q_2$. Now, for a spinning
particle moving in Minkowski space, $\alpha=1$, we obtain $p^2=p^3=0$, and hence ${\dot\theta}={\dot\phi}=0$. Thus, 
$rs^{12}=const.$, $r\sin\theta s^{13}=conts.$, as expected. If the particle has no spin, then these constants can simply be set equal to zero. 
For the case of motion of non-spinning particle in the curved space, we can still rescale the constants $q_{1,2,3}$ by a factor of $1-\alpha^2$ to 
have $r^2p^2=(1-\alpha^2)Q_1$, $r^2\sin\theta p^3=(1-\alpha^2)Q_2$. This ensures the desired behavior 
$p^{2,3}\rightarrow 0$ as $\alpha\rightarrow 1$. Now we study a few particular situations.


\subsection{Radial motion}\label{radial}

For radial motion, ${\dot\theta}={\dot\phi}=0$, Eqs. (\ref{e91f}) and (\ref{e91g}) result in
\begin{equation}
	p^2\,=\, p^3\,=\,0\label{e28}
\end{equation}
while $p^0$ and $p^1$ are given by Eqs. (\ref{g1}) and (\ref{g2}), respectively. Also, Eq. (\ref{e11}) gives $p^1=\beta$ where $\beta$
is defined in Eq. (\ref{beta1}). Putting this into Eq. (\ref{g2}) results in
\begin{equation}\label{e32}
	t = \frac{E}{m}\tau,
\end{equation}
which in turn, taking Eq. (\ref{e11a}) into account, gives
\begin{equation}
	{\dot r}=\frac{\beta}{m}.\label{e33}
\end{equation}

Regarding the spin components, from Eqs. (\ref{je2}) and (\ref{sp1})-(\ref{sp4}) we obtain
\begin{align}
	S^0\,&=\,\frac{Q_3\,p^1}{m\alpha}\,,\label{sp1a}\\
	S^1\,&=\,\frac{\alpha\,E\,Q_3}{m},\label{sp2a}\\
	S^2\,&=\,-\frac{m\,Q_2}{E\alpha r},\label{sp3a}\\
	S^3\,&=\,\frac{m\,Q_1}{E\alpha r\sin\theta}.\label{sp4a}
\end{align}
Note that $Q_{1,2,3}$ defined by Eqs. (\ref{k10a})-(\ref{k12a}) are constant for radial motion.
The particle spin orientation remains constant. In other words, the radial motion is possible if spin of the particle satisfies
Eqs. (\ref{sp1a})-(\ref{sp4a}). The above equations establish relations between the three constants $q_{1,2,3}$ and
the three spin components $S^{1,2,3}$.


\subsection{Motion in the equatorial plane}\label{planar}

For $\theta=\frac{\pi}{2}$, the geometry is that of a cone with deficit angle $(1-\alpha^2)\,\pi$.
Obviously we have, ${\dot\theta}=0$. As we stated earlier in Sec. \ref{who}, motion in this plane is possible if we have
$s^{23}=0$, by which, Eqs. (\ref{k12}) and (\ref{k12a}) give $q_2=q_3=0$. From Eqs. (\ref{k10a})-(\ref{k12a}), this leads to 
$Q_1=Q_3=0, Q_2=q_1$. Again, $p^0$ and $p^1$ are given by Eqs. (\ref{g1}) and (\ref{g2}). From Eq. (\ref{e91f}) we have
\begin{equation}
	p^2\,=\,0,\label{e38}
\end{equation}
and from Eq. (\ref{e11}) we obtain
\begin{equation}
	p^3\,=\,\pm\,\frac{\sqrt{\beta^2-(p^1)^2}}{r\,\alpha}.\label{e36}
\end{equation}
Equations (\ref{sp1})-(\ref{sp4}) result in
\begin{align}
	S^0\,&=\,S^1\,=\,S^3\,=\,0,\label{e36d}\\
	S^2\,&=\,\frac{m\,(r^2\,p^3-q_1)}{E\alpha r},\label{fo1}
\end{align}
which mean that the spin is perpendicular to the plane of motion. The trajectory can be obtained from 
Eqs. (\ref{e91g})-(\ref{e91e}), which reduce to
\begin{align}
	{\dot t}\,&=\,\frac{E}{m},\label{e36m}\\
	{\dot r}\,&=\,\frac{p^1}{m},\label{e36mm}\\
	{\dot\phi}\,&=\,\frac{p^3}{m}.\label{e36mmm}
\end{align}
The set of constants $q_1=q_2=q_3=0$ correspond to a non-spinning particle at rest.


\subsection{Geodesics}\label{geod}

A particle without spin will move along the geodesics of the spacetime. To recover the geodesics,
we set $s^{23}=0$ in Eq. (\ref{k12}), which in turn results in $Q_3=0$. Then, Eq. (\ref{k12a}) results in
\begin{equation}\label{ident}
	q_1\cot\theta\,=\,q_2\sin\phi+q_3\cos\phi.
\end{equation}
Also Eqs. (\ref{k10})-(\ref{k11a}) (after the rescaling discussed earlier in this section) together with Eq. (\ref{ident}) give
\begin{align}
	m\,r^2\,{\dot\theta}\,&=\,(1-\alpha^2)(-q_2\cos\phi + q_3\sin\phi),\label{ged1}\\
	m\,r^2\,\sin^2\theta\,{\dot\phi}\,&=\,(1-\alpha^2)q_1.\label{ged2}
\end{align}
Finally, from Eqs. (\ref{e9}), (\ref{u1}), (\ref{ged1}), (\ref{ged2}) we obtain
\begin{equation}\label{ged3}
{\dot r}\, = \pm\frac{\sqrt{\beta^2 r^2-\alpha^2(1-\alpha^2)^2 q^2}}{m r} 
\end{equation}
which can be solved to give
\begin{equation}\label{ged300}
r^2\, =\, \rho^2_0+\left(\frac{\beta\tau}{m}\pm\sqrt{r^2_0-\rho^2_0}\right)^2,
\end{equation}
where $\rho_0\equiv\frac{\alpha(1-\alpha^2)q}{\beta}$, and the $\pm$ sign corresponds to the $\pm$ sign in Eq. (\ref{ged3}). 
By inserting Eq. (\ref{ged300}) into Eqs. (\ref{ged1}) and(\ref{ged2}), and integrating the resulting system of equations,
one can obtain the trajectory of the particle. In this case, $q_{1,2,3}$ are related to the
initial values of $p^{1,2,3}$ or, equivalently, the initial values of ${\dot r}, {\dot\theta}, {\dot\phi}$. 


\section{Non-geodesic motion}\label{nogeodesy}
Now, we return to the general situation described in Sec. \ref{genmot}. We first need to determine parameters of the problem.
The parameter $\alpha$ is specified by the geometry of spacetime. The particle has two characteristic parameters $m$ (mass) and 
$s$ (spin), and starts at a specific point described by three parameters $r_0,\theta_0, \phi_0$. Then, we have a set of first order differential equations for 
the motion of the particle, so we need one extra parameter for each equation. These include four equations for the spin four-vector. 
The corresponding parameters are given by $q_{1,2,3}$. The fourth spin-related parameter is not independent, and is given in terms of 
$q_{1,2,3}$ through Eq. (\ref{e8aa}). Next, we have four equations for the four-momentum components. The first momentum-related 
parameter is $E$, the particle's energy. The other three momentum-related parameters are not independent, and are given in terms of the 
other parameters through Eqs. (\ref{e9}), (\ref{k10}), and (\ref{k11}). The trajectory of the particle is obtained by integrating 
the components of the four-momentum. Therefore, we need four trajectory-related parameters, but we already have the three parameters 
$(r_0,\theta_0,\phi_0)$. The fourth trajectory-related parameter is not independent, and is fixed by Eq. (\ref{e11a}). 
We also need an expression relating $q$, defined in Eq. (\ref{qqq}) 
to the spin of the particle, $s$. Such a relation is established by Eq. (\ref{e61}), from which we obtain 
\begin{align}
	r_0^2\, &=\, \left(\frac{1-\alpha^2}{m\lambda}\right)^2\left\{\left(\frac{Es\alpha}{m}\right)^2- \alpha^2 q^2
	\right.\nonumber\\&\quad +\left.(\alpha^2-\alpha^4-\lambda^2){\mathcal Q}_0\right\},	 \label{e6100}
\end{align}
in which ${\mathcal Q}_0\equiv{\mathcal Q}(\theta=\theta_0,\phi=\phi_0)$. Thus, by using Eq. (\ref{e6100}), we can rewrite 
Eq. (\ref{e61}) as 
\begin{equation}
	r^2\, =\, r_0^2+\frac{(1-\alpha^2)^2}{m^2\lambda^2}(\alpha^2-\alpha^4-\lambda^2)({\mathcal Q}-{\mathcal Q}_0).\label{e6200}
\end{equation}

Finally, the constant $\lambda$ (introduced by Eqs. (\ref{ansz7})-(\ref{ansz9})) can be obtained in terms of the other parameters. 
By inserting Eq. (\ref{k20a}) into Eq. (\ref{k200}), and making use of Eqs. (\ref{u1}) and (\ref{ansz9}), we obtain 
\begin{equation}\label{siginit}
	\lambda^2=\frac{\gamma^2(q^2-{\mathcal Q}_0)(\beta^2r^2_0-\alpha^2\gamma^2{\mathcal Q}_0)}
	{m^4 r^4_0+\gamma{\mathcal Q}_0(\alpha^2\gamma q^2+\gamma^2{\mathcal Q}_0+2m^2r^2_0)},
\end{equation}
in which $\gamma\equiv 1-\alpha^2$.

\begin{figure}[h]
	\centering
	\includegraphics[width=0.46\textwidth]{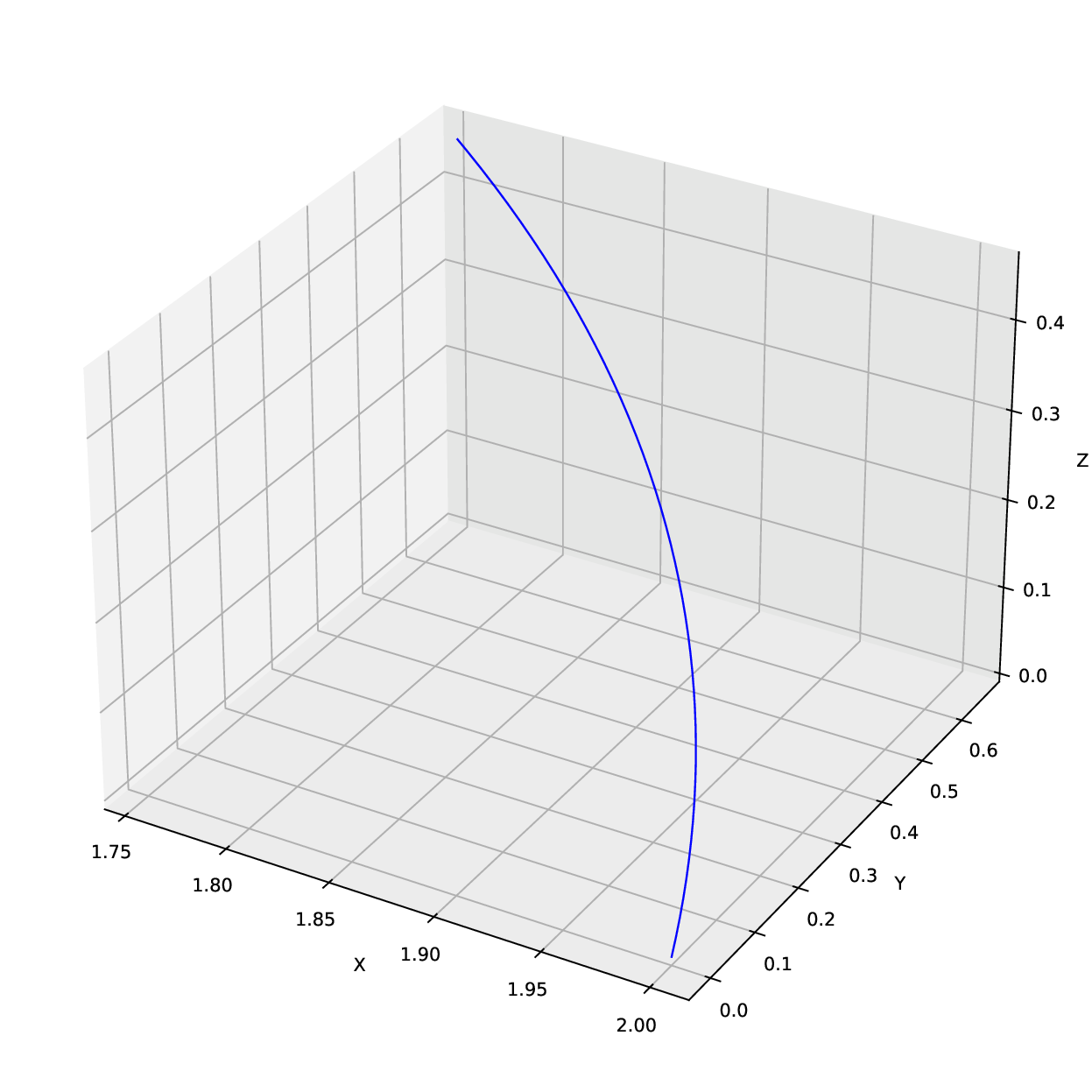}
	\caption{Trajectory of the spinning particle for $\alpha=0.95,\theta_0=\frac{\pi}{2},\phi_0=0,r_0=2,m=1,E=1.01,q_1=q_2=q_3=0.1, 
	0\leq\tau\leq 120$.}
	\label{plot0}
\end{figure}

Now, we can insert the above data into Eqs. (\ref{pl1}) and (\ref{pl2}) which can then be integrated to give 
$r(\tau), \theta(\tau), \phi(\tau)$. Due to the symmetries of the spacetime we can set $\phi_0=0$. We also set $m=1$. 
The trajectory of the particle is shown in Fig.\ref{plot0} for $\theta_0=\frac{\pi}{2}$, $E=1.01$, $q_1=q_2=q_3=0.1$, $\alpha=0.95$,
$r_0=2$. Then, Eq. (\ref{siginit}) gives $\lambda\approx 0.0007$. This set of parameters corresponds, according to Eq. (\ref{e6100}), to a 
particle with spin $s\approx 0.17$. 

\begin{figure}[h]
	\centering
	\includegraphics[width=0.46\textwidth]{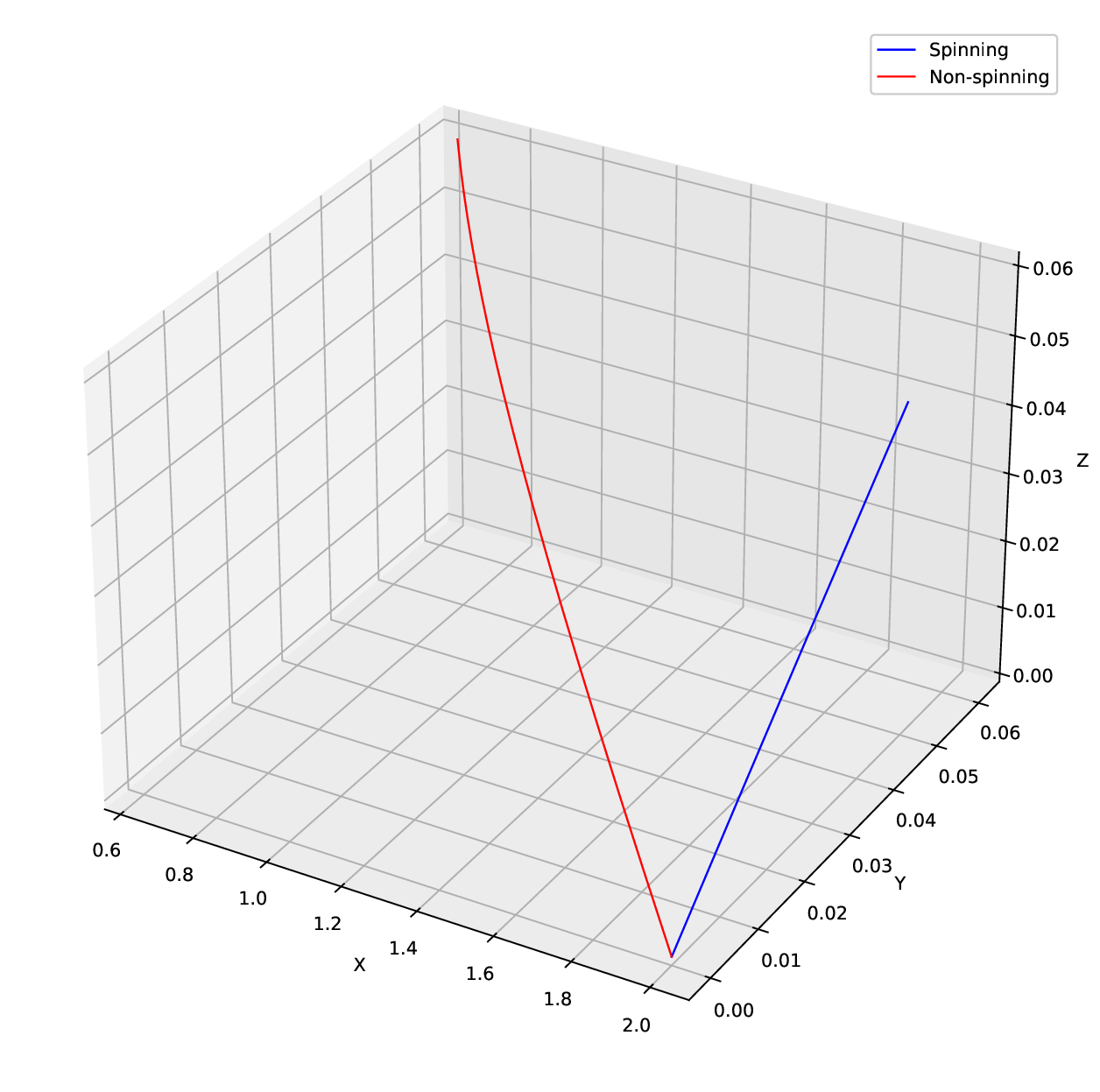}
	\caption{Trajectory of the particles for $\alpha=0.95,\theta_0=\frac{\pi}{2},\phi_0=0,r_0=2,m=1,E=1.01,q_1=q_2=0.1,0\leq\tau\leq 10$.
	For the spinning particle, $q_3=0.1$, and for the corresponding geodesic, $q_3=0$.}
	\label{plot1}
\end{figure}

To show the effect of spin-curvature coupling on the trajectory of the particle, trajectories of a spinning particle and a 
corresponding geodesic are shown in the same plot in Fig.\ref{plot1}. For the spinning particles, the parameters are chosen as in 
Fig.\ref{plot0}, and the geodesic is basically obtained by inserting Eq. (\ref{ged300}) (with the minus sign) into Eqs. (\ref{ged1}) 
and (\ref{ged2}) and integrating the resulting equations with the initial condition $(\theta,\phi)(0)=(\theta_0,\phi_0)$. To determine 
the relevant parameters, we first insert $\theta_0=\frac{\pi}{2}, \phi_0=0$ into Eq. (\ref{ident}), which gives $q_3=0$. We also require 
the non-spinning particle moving along this geodesic to have the same initial velocity components ${\dot\theta}(0), {\dot\phi}(0)$ as the 
spinning particle at the point $(r_0,\theta_0,\phi_0)$. To satisfy this requirement, we compare the pair of equations (\ref{ged1}), 
({\ref{ged2}}) with the pair of equations (\ref{pl1}) and (\ref{pl2}), to obtain $q_1\approx 0.10, q_2\approx 0.1$. Finally, the energy of the 
non-spinning particle can be obtained by requiring that both particles have the same velocity components ${\dot r}(0)$. Then, comparing 
Eq. (\ref{ged3}) with the pair of equations (\ref{e91e}) and (\ref{k20a}) yields $E\approx 1.01$. Figure \ref{plot1} illustrates that a 
spinning particle and a non-spinning particle, both starting from the same initial point with identical polar and azimuthal momenta, deviate 
from each other, and this deviation increases over time. For the values of the parameters chosen here, the non-spinning particle moves rather 
rapidly inwards, while the spinning particle trajectory deviates only slightly toward $r<r_0$ as a result of the spin-curvature coupling.   

\begin{figure}[h]
 	\centering
 	\includegraphics[width=0.46\textwidth]{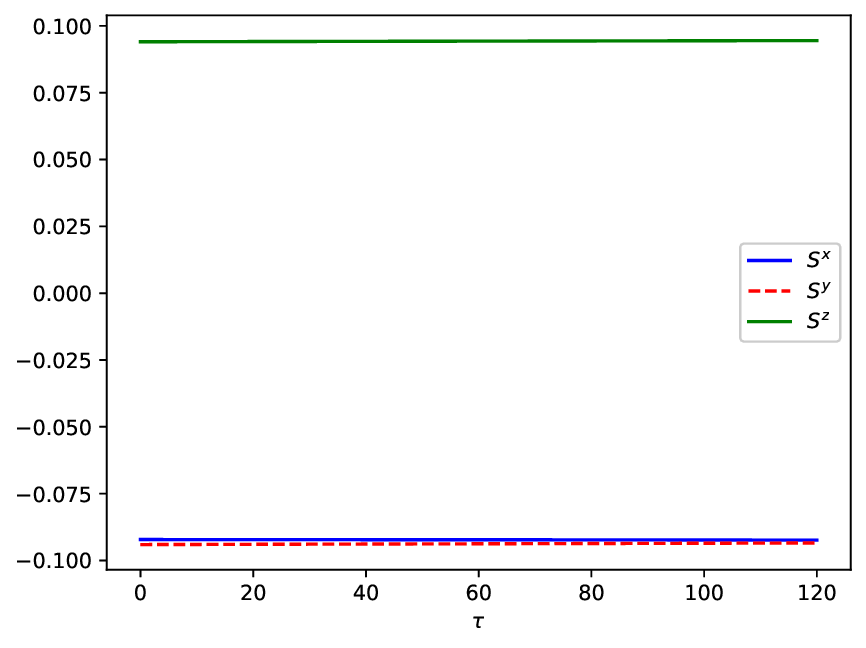}
 	\caption{The evolution of the components of the spin along the trajectory of the particle for
 		$\alpha=0.95, E=1.01, q_1=q_2=q_3=0.1,\theta_0=\frac{\pi}{2},\phi_0=0,r_0=2,m=1$. }
		\label{plot2}
\end{figure}

\begin{figure}[h]
 	\centering
 	\includegraphics[width=0.46\textwidth]{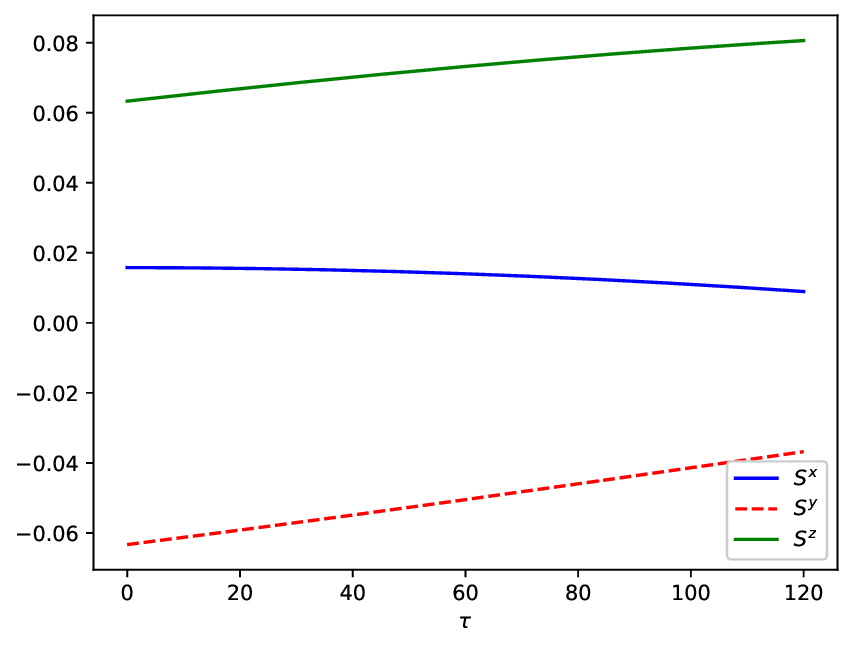}
 	\caption{The evolution of the components of the spin along the trajectory of the particle for
 		$\alpha=0.95, E=1.5, q_1=q_2=q_3=0.1,\theta_0=\frac{\pi}{2},\phi_0=0,r_0=2,m=1$. }
		\label{plot5}
\end{figure}

The evolution of the components of spin
\begin{align}
S^x&=(\sin\theta S^1+r\cos\theta S^2)\cos\phi-r\sin\phi\sin\theta S^3,\label{xyz1}\\
S^y&=(\sin\theta S^1+r\cos\theta S^2)\sin\phi+r\cos\phi\sin\theta S^3,\label{xyz2}\\
S^z&=\cos\theta S^1-r\sin\theta S^2,\label{xyz3}
\end{align}
are shown in Fig.\ref{plot2} representing almost constant overall orientation of the spin vector. Note that for the chosen parameters,
the $S^x$ and $S^y$ components almost coincide. For other choices of the parameters, say larger values of $E$, three distinct and 
slightly-changing curves are obtained. This is shown in Fig.\ref{plot5} for $E=1.5$ (and the rest of the parameters as above) 
which corresponds to $\lambda\approx 0.005$ and $s\approx 0.13$.      

\begin{figure}[h]
          \vspace{3mm}
	\centering
	\includegraphics[width=0.46\textwidth]{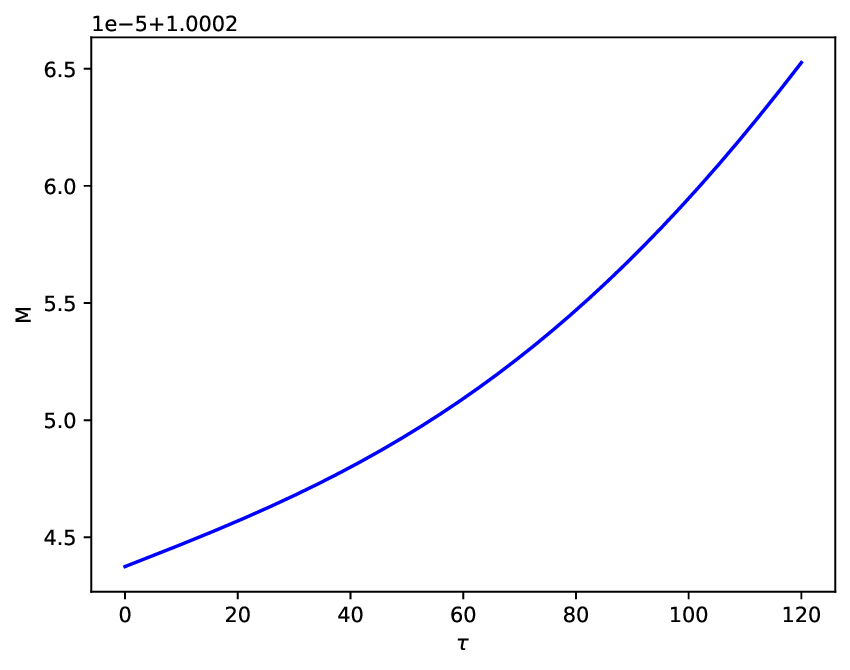}
	\caption{The evolution of $M$ along the trajectory of the particle for
		$\alpha=0.95, E=1.01, q_1=q_2=q_3=0.1,\theta_0=\frac{\pi}{2},\phi_0=0,r_0=2,m=1$.}
	\label{plot3}
\end{figure}

The behavior of $M$, defined by Eq. (\ref{e912d}), is shown in Fig.\ref{plot3}. The small increase of the value of $M$ can be 
qualitatively explained as follows: as Fig.\ref{plot0} shows, as the spinning particle moves along its trajectory, its distance to the 
origin, $r$, decreases gradually, and hence the second term in the right-hand side of Eqs. (\ref{e912d}), which is proportional to $r^{-2}$, 
increases with time by a small amount. For longer time intervals, $M$ reaches a local maximum value and then decreases for some time. 
 
\section{Discussion}
We studied the motion of spinning test particles in the background spacetime of a global monopole. By using the symmetries
of the spacetime, we found an exact solution to the MPD equations. To the best of our knowledge, this is the first general exact 
solution to the equations of motion of a spinning test particle moving in a curved spacetime background within the MPD framework.
We also showed that the system is completely integrable by computing the relevant Poisson brackets.
We investigated the particle's trajectories, momenta and spins for certain values of the underlying parameters and compared it 
with a corresponding geodesic. We also examined the particular cases of radial and planar motion. Although the exact solution we
presented in this work has its own merits, at least from a formal point of view, it is also of use for a better understanding
of the global monopole spacetime, given the fact that the global monopole spacetime has received rather limited attention so far.
In particular, the present work reveals several interesting properties of the global monopole spacetime. It
accommodates both geodesic and non-geodesic motion of spinning particles. It represents a curved spacetime in which the spin-curvature 
coupling is nonvanishing, yet the geometry is sufficiently symmetric to allow an exact solution of the highly coupled MPD equations. 
It also provides an explicit example of the importance of the hidden symmetries of spacetime, encoded in the Killing-Yano tensor, 
for the complete integrability of the equations of motion. In fact, the properties of the spacetime allow us to obtain an exact 
conserved quantity while retaining the quadratic spin terms, whereas these terms usually have to be neglected in other spacetimes 
admitting Killing-Yano tensors.

This study might be extended by investigating more exact solutions, if any, and by considering motion of particles with a 
richer internal structure, namely those with both spin and quadrupole moments. It might also be interesting to see if a similar 
analysis is possible for other spacetimes with topological defects.

\acknowledgments
M.M. would like to thank the Department of Physics and Astronomy, University of Lethbridge for hospitality.
We thank the Natural Sciences and Engineering Research Council of Canada under Grant No. RGPIN-2019-05404 for support. We would like to thank
an anonymous referee of Physical Review D for invaluable comments and suggestions.

\bibliographystyle{apsrev4-2}
\bibliography{monoabrev3}

\end{document}